\documentclass[journal]{IEEEtran}

\usepackage{cite}
\usepackage{amsmath,amssymb,amsfonts}
\usepackage{algorithmic}
\usepackage{graphicx}
\usepackage{textcomp}
\usepackage{adjustbox}
\usepackage{xcolor}
\def\BibTeX{{\rm B\kern-.05em{\sc i\kern-.025em b}\kern-.08em
    T\kern-.1667em\lower.7ex\hbox{E}\kern-.125emX}}
    
\begin{document}

\IEEEoverridecommandlockouts
\IEEEpubid{\begin{minipage}{\textwidth}\ \\[20pt]
\centering\normalsize{\copyright 2021 IEEE. DOI: 10.1109/ACCESS.2021.3096853}
\end{minipage}}

\title{Deep Learning Based  Carrier Frequency Offset Estimation in IEEE 802.11ah\\
}



\author{Vukan Ninkovic,~\IEEEmembership{Student Member,~IEEE,}
        Aleksandar Valka,
        Dejan Dumic,
        and Dejan Vukobratovic,~\IEEEmembership{Senior Member,~IEEE}
\thanks{V. Ninkovic and D. Vukobratovic are with the Department of Power, Electronics and 
Communications Engineering, University of Novi Sad, 21000, Novi Sad,
Serbia (e-mail: ninkovic@uns.ac.rs; dejanv@uns.ac.rs)
.}
\thanks{A. Valka and D. Dumic are with the Methods2Business, Mite Ruzica 1, 21000, Novi Sad,
Serbia (e-mail: aleksandar@methods2business.com; dejan@methods2business.com).}
}

\maketitle

\begin{abstract}
Wi-Fi systems based on the IEEE 802.11 standards are the most popular wireless interfaces that use Listen Before Talk (LBT) method for channel access. The distinctive feature of a majority of LBT-based systems is that the transmitters use preambles that precede the data to allow the receivers to perform packet detection and carrier frequency offset (CFO) estimation. Preambles usually contain repetitions of training symbols with good correlation properties, while conventional digital receivers apply correlation-based methods for both packet detection and CFO estimation. However, in recent years, data-based machine learning methods are disrupting physical layer research. Promising results have been presented, in particular, in the domain of deep learning (DL)-based channel estimation. In this paper, we present a performance and complexity analysis of packet detection and CFO estimation using both the conventional and the DL-based approaches. The goal of the study is to investigate under which conditions the performance of the DL-based methods approach or even surpass the conventional methods, but also, under which conditions their performance is inferior. Focusing on the emerging IEEE 802.11ah standard, our investigation uses both the standard-based simulated environment, and a real-world testbed based on Software Defined Radios.  

\end{abstract}


\section{Introduction}
\label{sec:introduction}
\IEEEPARstart{W}{ireless} communication systems based on the orthogonal frequency division multiplexing (OFDM) dominate current wireless research and development. In order to ensure fairness, wireless systems operating in unlicencsed bands share a common channel using Listen Before Talk (LBT) methodology. Common approach in a majority of LBT systems is that the transmitters send preambles prepended to data packets in order to ensure that the receivers detect signal and acquire initial synchronization. Preambles usually contain a sequence of symbols with good correlation properties, allowing the receiving end to identify packet start samples and establish initial timing and frequency offset synchronization. Conventional model-based signal processing methods at OFDM receivers are well understood and are currently used as a basis for the receiver design \cite{b1,b2,b3,b4,b5,b7,b9,b10,schlegel_2006}. 

Conventional methods are recently challenged by the data-based approaches relying on deep learning (DL) \cite{Wang_2017,OShea_2017,Qin_2019}. DL-based methods have been evaluated across the physical layer (PHY), ranging across signal detection \cite{Karra_2017}, channel estimation \cite{Ye_2018,He_2018} and error correction coding \cite{Nachmani_2018}, demonstrating promising performance as compared to the conventional methods. Moreover, the DL-based positioning services that exploit channel state information as fingerprints have been explored recently \cite{Wang_2016}. However, in most of the DL-based PHY studies, signal detection at the receiver that includes procedures that precede channel estimation, such as packet detection and carrier frequency offset (CFO) estimation, are assumed to be perfectly known. In addition, studies on DL-based PHY methods focusing specifically on preamble-based LBT OFDM systems are also missing, with an exception in the domain of channel estimation \cite{Han_2019}.

In this paper, we fill this gap by focusing on the DL-based methods for packet detection and CFO estimation in IEEE 802.11 systems. In order to provide a detailed, standard-specific investigation, we consider an emerging IEEE 802.11ah standard for low-power Internet of Things (IoT) applications \cite{b6}. We use both the standard-based simulated environment, and a real-world testbed based on  Software Defined Radios (SDRs) to evaluate our results.

The paper is topically divided in two parts. In the first part of the paper, we focus on the packet detection problem and provide a detailed complexity vs performance evaluation and comparison between the conventional and the DL-based packet detection. Our results demonstrate that the DL methods based on the one-dimensional Convolutional Neural Networks (1D-CNN) may outperform conventional methods under reduced computational effort, while being inferior in miss detection and false alarm rates.

In the second part of the paper, we investigate the DL-based CFO estimation methods and compare them to the conventional methods. Our results show that, for the CFO estimation at the IEEE 802.11ah receiver, long short-term memory (LSTM)-based recurrent neural network (RNN) are able to match the performance of the conventional methods, and even surpass them at low-to-medium signal-to-noise ratios (SNR). However, despite their excellent accuracy, DL-based methods suffer from higher complexity as compared to the conventional methods.

Our goal in this paper is to discuss both the benefits and drawbacks of DL-based methods in the context of a specific wireless standard (IEEE 802.11ah) and provide fair comparison with the conventional methods. In other words, the main message of the paper is not in advocating the usage of DL-based solutions, but in pointing out, in a given scenario, when it is advantageous to use such methods and when it is not. 

\subsection{Related Work and Paper Contributions}

Using DL for PHY processing is a very active research area. However, most of the recent work is focused on the channel estimation, assuming that the signal detection and synchronization is ideal. Nevertheless, several recent papers address the DL-based signal detection in several scenarios. 

Authors in \cite{Li2018} address the problem of CFO in the uplink of the OFDM access (OFDMA) system, where DL is used to suboptimally estimate CFOs corresponding to different users. The DL-based CFO for the received signals after a low resolution analog-to-digital conversion in emerging mmWave multiple-input multiple-output (MIMO) systems is investigated in \cite{Dreifuerst2020}, demonstrating improved performance as compared to the conventional methods. For OFDM-based unmanned aerial vehicle communications, DL methods for CFO are proposed in \cite{Kumari2021}. Our work on the CFO estimation part is influenced by \cite{Karra_2017}, an early study on DL-based CFO estimation in single-carrier systems. Finally, a comprehensive overview of DL methods for the IEEE 802.11ax receiver design is presented in \cite{Zhang2021}.

The contributions of this work are summarized as follows:
\begin{itemize}
    \item We introduce a DL-based packet detection in preamble-based IEEE 802.11 systems and provide systematic performance and complexity comparison with the conventional packet detectors. The initial results, presented in \cite{Ninkovic_2020}, are here expanded with additional numerical results and SDR-based real-world demonstrations;
    \item We present a systematic performance and complexity comparison of the DL-based and the conventional CFO methods in preamble-based IEEE 802.11 systems;
    \item Our results are demonstrated using standard-based IEEE 802.11ah simulated environment and verified in a real-world setup using SDRs;
    \item The study provides clear insights under which conditions the performance of the DL-based methods may approach or even surpass the conventional methods for packet detection and CFO estimation, but also, under which conditions their performance  is inferior.
\end{itemize}

To summarize, compared to  \cite{Ninkovic_2020}, this paper extends our work to a more challenging problem of CFO estimation, provides extensive simulation and SDR-based real-world performance results, and presents a detailed discussion on implementation complexity for both packet detection and CFO estimation.

The paper is organized as follows. In Sec. \ref{Sec2}, we present a system model and review IEEE 802.11ah frame structure. Sec. III deals with the packet detection problem, where the conventional and the DL-based methods are first described, and then evaluated using numerical simulations and the real-world SDR experiments. In a similar manner, Sec. IV describes and compares the conventional and the DL-based CFO estimation methods, including simulated and real-world SDR-based results. The paper is concluded in Sec. V.

\section{Background and System Model}
\label{Sec2}

\subsection{OFDM Communication System Model}

We consider a conventional OFDM system with $N$ subcarriers separated by $\Delta f$ in the frequency domain. At the transmitter side, the binary information sequence is mapped onto the sequence of complex modulation symbols $\pmb{X}$ allocated to different subcarriers and converted into the time-domain signal $\pmb{x}$ via Inverse Discrete Fourier Transform (IDFT) \cite{b1}. The resulting discrete-time complex baseband signal is obtained as:
\begin{equation}
\begin{split}
x_n= \frac{1}{N} \sum_{k=0}^{N-1} \ X_k e^\frac{j (2 \pi k n)}{N}, \quad n = 0, 1, ..., N-1    , 
\end{split}
\end{equation}
where $X_k$ are the complex samples in the frequency domain.

Cyclic prefix (CP) of length greater than the expected channel delay spread is inserted in order to mitigate Inter-Symbol Interference (ISI) and preserve the orthogonality of the subcarriers \cite{b2}. After oversampling and filtering, the oversampled signal $\pmb{x}_{os}$ will propagate through the indoor multipath channel. Focusing on the discrete-time complex-baseband model, the channel is represented via an equivalent discrete-time impulse response $\pmb{h}$. After the complex additive white Gaussian noise (AWGN) $\pmb{w}$ samples are added, the discrete-time complex-baseband signal at the receiver side can be obtained as: 
\begin{equation}
\pmb{y}_{os} = \pmb{x}_{os} \circledast \pmb{h} + \pmb{w},
\end{equation}
where $\circledast$ represents the circular convolution.

\begin{figure}[htbp]
 \centerline{\includegraphics[width=1\columnwidth]{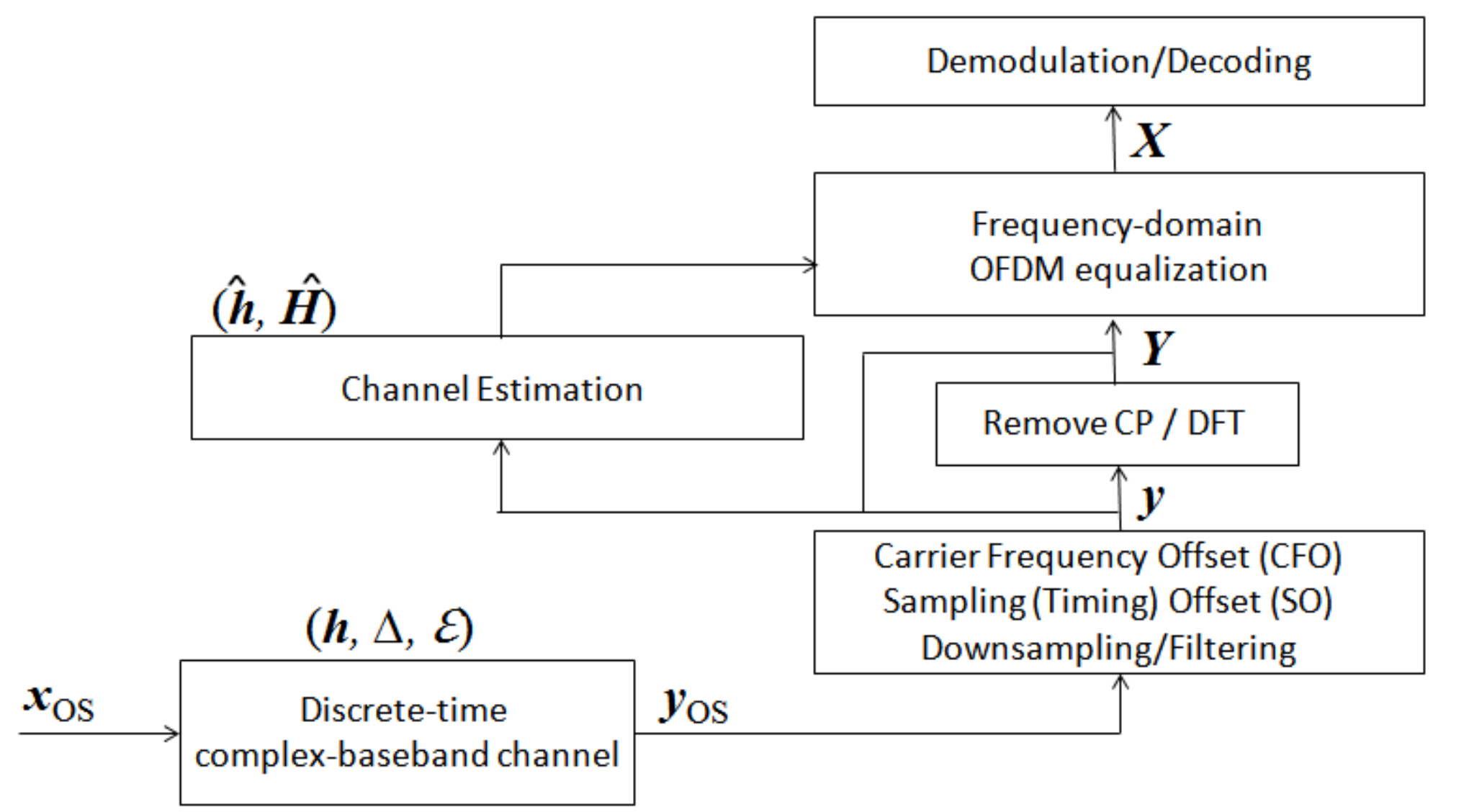}}
  \caption{Generic Architecture of the OFDM Wireless Receiver.}
  \label{fig:OFDM_system_model}
\end{figure}  

The receiver side, which is in the focus of this paper, is illustrated in Fig. \ref{fig:OFDM_system_model}. After the signal passes through a reverse pulse shaping filter, it is  downsampled, time and frequency offsets are corrected, and the cyclic prefix is removed. In order to demodulate the received signal, DFT is performed, and the frequency-domain signal is written as:
\begin{equation}
Y_k = \sum_{n=0}^{N-1} \ y_n \cdot e^{-\frac{j(2\pi k n)}{N}}, \quad k = 0, 1, ..., N-1 
\end{equation}
Next, the signal correction using channel estimation techniques (usually based on inserted pilot symbols) is executed and the data is passed to the signal demapper block for the demodulation and channel decoding. Lastly, the binary information data is obtained back.

Note that, besides the channel impairment and the noise, the received signal ($\pmb{y}_{os}$) is affected by the time sampling offset ($\varepsilon = \frac{\tau_{off}}{T}$, where T represents duration of one OFDM symbol) and the carrier frequency offset ($\Delta = \frac{f_{off}}{\Delta f}$), which needs to be estimated and corrected. A carrier frequency offset (CFO) of $f_{off}$ causes a phase rotation of $2\pi tf_{off}$. If uncorrected, this causes both a rotation of the constellation and a spread of the constellation points similar to the AWGN. A timing error will have a little effect as long as all the taken samples are within the length of the cyclically-extended OFDM symbol \cite{b3}.

\subsection{IEEE 802.11ah Frame Structure} 

\begin{figure}[t]
 \centerline{\includegraphics[width=1\columnwidth]{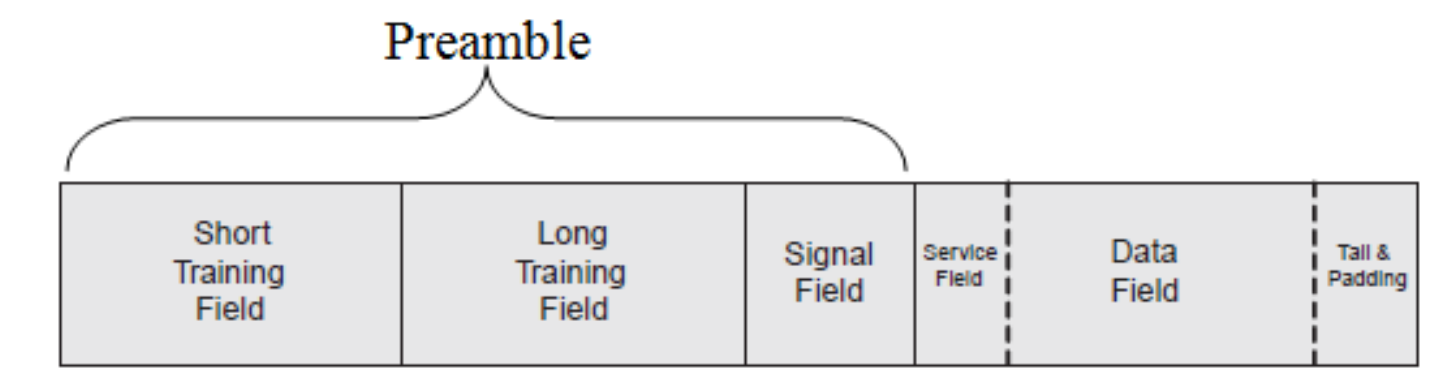}}
  \caption{IEEE 802.11 preamble and frame structure.}
  \label{fig:OFDM_frame_structure}
\end{figure}

In this paper, we focus on listen-before-talk (LBT)-based IEEE 802.11 OFDM technologies, whose frame structure is shown in Fig. \ref{fig:OFDM_frame_structure}. In LBT systems, the sequence of data symbols is preceded by a preamble of known data needed to establish the initial synchronization and/or channel estimation \cite{b4}. The initial synchronization includes the frame detection (estimation of the initial time sample of the incoming frame) and frequency offset estimation. Preamble structure is usually based on a certain repeated pattern, representing sequences with good correlation properties that provide for good time and frequency synchronization \cite{b5}. 

For the purpose of detailed implementation and evaluation, in this paper, we restrict our attention to the IEEE 802.11ah (\textit{Wi-Fi HaLow}) standard. The 802.11ah 1 MHz packet preamble is a pilot sequence with a fixed length of 14 OFDM symbols (for single-antenna transmission) where each OFDM symbol has $N=32$ subcarriers of subcarrier spacing $\Delta f = 31.25~kHz$. Normal cyclic prefix of $8~\mu s$ duration is applied, resulting in $40~\mu s$ OFDM symbol \cite{b7}. Note that the composition of the preamble remains the same as in conventional 802.11 systems, further adapted to specific 802.11ah requirements \cite{b4,b6,b7}:  

\textit{Short Training Field } (STF) - The short training field, which lasts $160~\mu s$, consists of 4 OFDM symbols in the frequency domain which, after IDFT, represent 10 repetitions of the same short training symbol ($16~\mu s$ each) in the time domain. Short training symbol is a sequence with good correlation properties and a low peak-to-average power whose features are preserved even after clipping or compression by an overloaded analog front end. Because of that, a short training field is suitable for coarse timing synchronization (packet detection) and (coarse) frequency offset estimation.

\textit{Long Training Field 1} (LTF1) - The first long training field also contains 4 OFDM symbols of $160~\mu s$ duration. Two repetitions of the same long training symbol enables fine timing synchronization, fine frequency offset estimation and channel estimation.

\textit{Signal Field} (SIG) - The signal field, which is made of 6 OFDM symbols, contains packet information to configure the receiver: rate (modulation and coding), length (amount of data being transmitted in octets), etc.

\textit{Long Training Field 2} (LTF2) - The second long training field is used for MIMO channel estimation, and in our case, because only SISO transmission is applied, this part does not exist.


\begin{figure}[htbp]
 \centerline{\includegraphics[width=1\columnwidth]{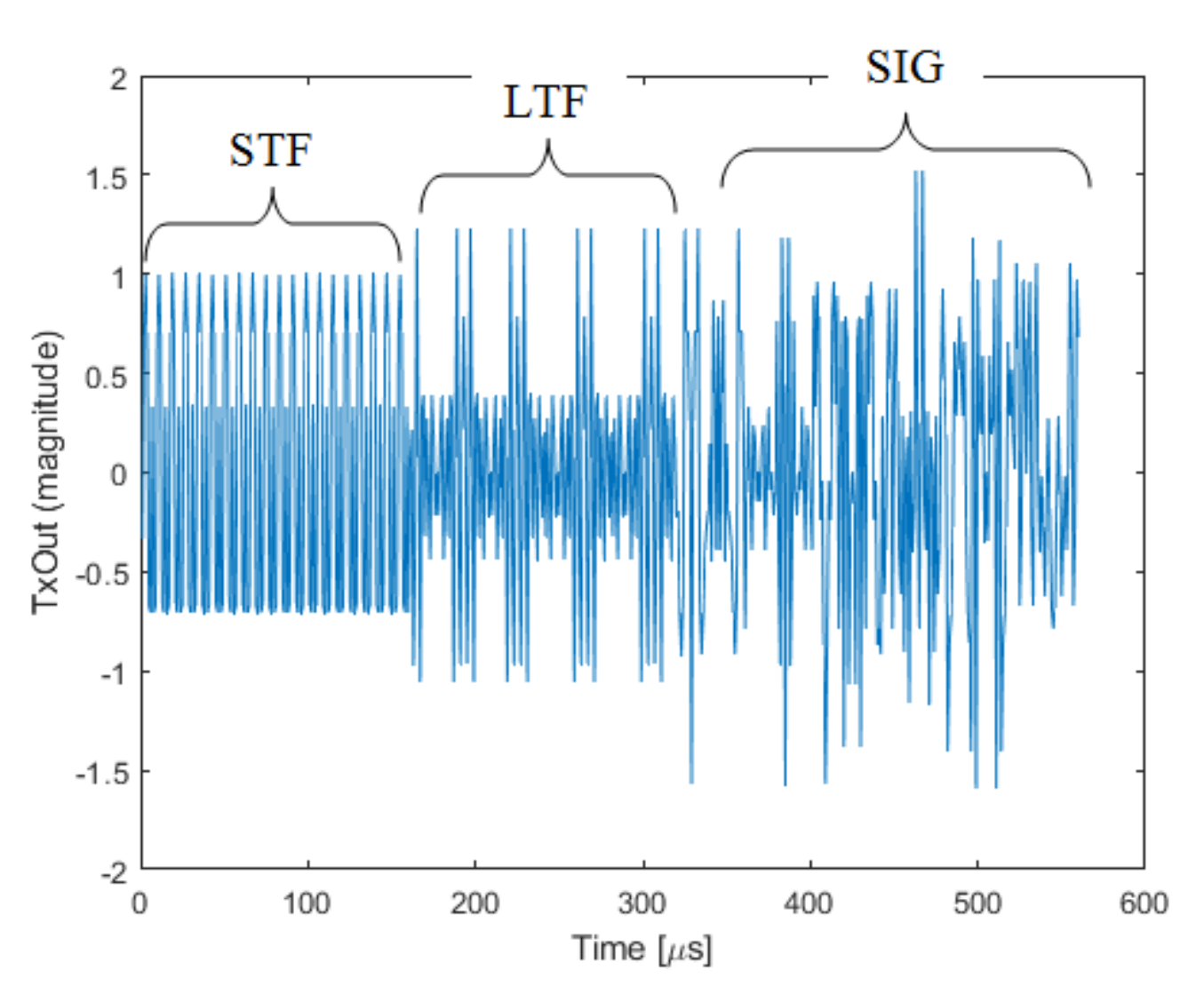}}
  \caption{802.11ah packet preamble -- the transmitted waveform.}
  \label{fig:NDP_packet}
\end{figure}

In this paper, we focus on the problem of initial synchronization, which depends only on the packet preamble. To reduce the complexity of both the simulations and real-world experiments, 802.11ah Null Data Packet (NDP) \cite{b8} is used, containing only the preamble (without data field). The transmit waveform of the NDP packet is shown in Fig. \ref{fig:NDP_packet}.

\section{Preamble-Based Packet Detection}

\subsection{Conventional Packet Detection Methods}
\label{Conventional_packet_det}

Conventional algorithms for packet detection, which are nowadays widely used, use repetitive preamble structure through complex correlation between two subsequently received training symbols. If we suppose that the number of complex samples in one training symbol is $L$, such complex correlation can be expressed as:

\begin{equation}
\Lambda_{\tau} = \sum_{i=0}^{L-1} \ y^{*}_{\tau+i} y_{\tau+i+L}
\end{equation}

 In \cite{b3} and \cite{b10}, the authors proposed a packet detection algorithm which relies on the assumption that the channel effects will be annulled if the conjugated sample from one training symbol is multiplied by the corresponding sample from the adjacent training symbol. Consequently, products of these sample pairs at the start of the frame will have approximately the same phase, thus the magnitude of their sum will be a large value. In order to reduce the complexity of the algorithm, they introduced a window of $2L$ samples which slides along the time $\tau$ as the receiver searches for the first training symbol, i.e., the packet start sample $\tau_S$. Timing metric used for the packet detection is: 

\begin{equation}
M(\tau) = \frac{|\Lambda_{\tau}|^2}{P_{\tau}^2},
\end{equation}
where $P_{\tau}$ is the sum of the powers of $L$ subsequent samples:

\begin{equation}
P_{\tau} = \sum_{i=0}^{L-1} \ |y_{\tau+i+L}|^2
\end{equation}

From the timing metric $M(\tau)$, one may find the initial packet sample by finding the sample that maximizes $M(\tau)$. In addition, except finding the maximum sample-point, observing the points to the left and right in the time domain which are at the 90\%\ of the maximum, and averaging these two $90\%$-time samples, may result in more accurate timing estimation. A threshold which triggers the above algorithm should be chosen in a way that the algorithm minimizes the probability of miss detection while controlling for the probability of false alarm.

Packet detection in IEEE 802.11 is usually separated into two steps: coarse and fine synchronization, where the main principles from conventional algorithms are reused and adapted to the specific system requirements. The coarse packet detection, denoted as $\hat{\tau}_{S}$, may follow \cite{b3} (Eq. 5), setting $L=80$ samples (one half of the STF duration):

\begin{equation}
\begin{gathered}
    \hat{\tau}_{S}=\arg\max_{\tau}\frac{|\Lambda_{\tau}|^2}{(P_{\tau})^2}\\
    =\arg\max_{\tau}(\frac{|\sum_{n=\tau}^{\tau+L_{S}-l_{S}} y^*_{n} y_{n+l_{S}}|^2}{(\sum_{n=\tau}^{\tau+L_{S}-l_{S}} |y_{n+l_{S}}|^2)^2}),
\end{gathered} 
\end{equation}
where $l_{S}$ is the STS sample-length and $L_{S}$ represents the sample-lengths of the STF field. After calculating $\hat{\tau}_{S}$, we can extract the whole preamble because the peaks from the correlation between a single long training symbol and the entire preamble are used to derive more accurate time estimation \cite{b4}. 

\subsection{Deep-Learning Based Packet Detection}
\label{DL_packet_det}

The packet detection problem can be formulated as a regression problem, where DNN needs to learn a mapping between the input signal and the output value representing the packet start instant while distinguishing from the noise. We suppose that DNN-based packet detection operates over the consecutive fixed-length blocks $|\pmb{y}|$ of the received signal amplitude samples:

\begin{equation}
    \hat{\tau_{S}} = f(|\pmb{y}|),
\end{equation}
after the received signal is downsampled and filtered. Next, we will describe the DNN architecture used for packet detection task, as well as the training procedure. 

\subsubsection{Convolutional Neural Networks for Packet Detection} Motivated by recent investigation in \cite{Karra_2017} and the initial results obtained in \cite{Ninkovic_2020}, we consider Wi-Fi packet detection using one-dimensional convolutional neural networks (1D-CNN).

\begin{figure}[htbp]
 \centerline{\includegraphics[width=3.2in, height=2.4in]{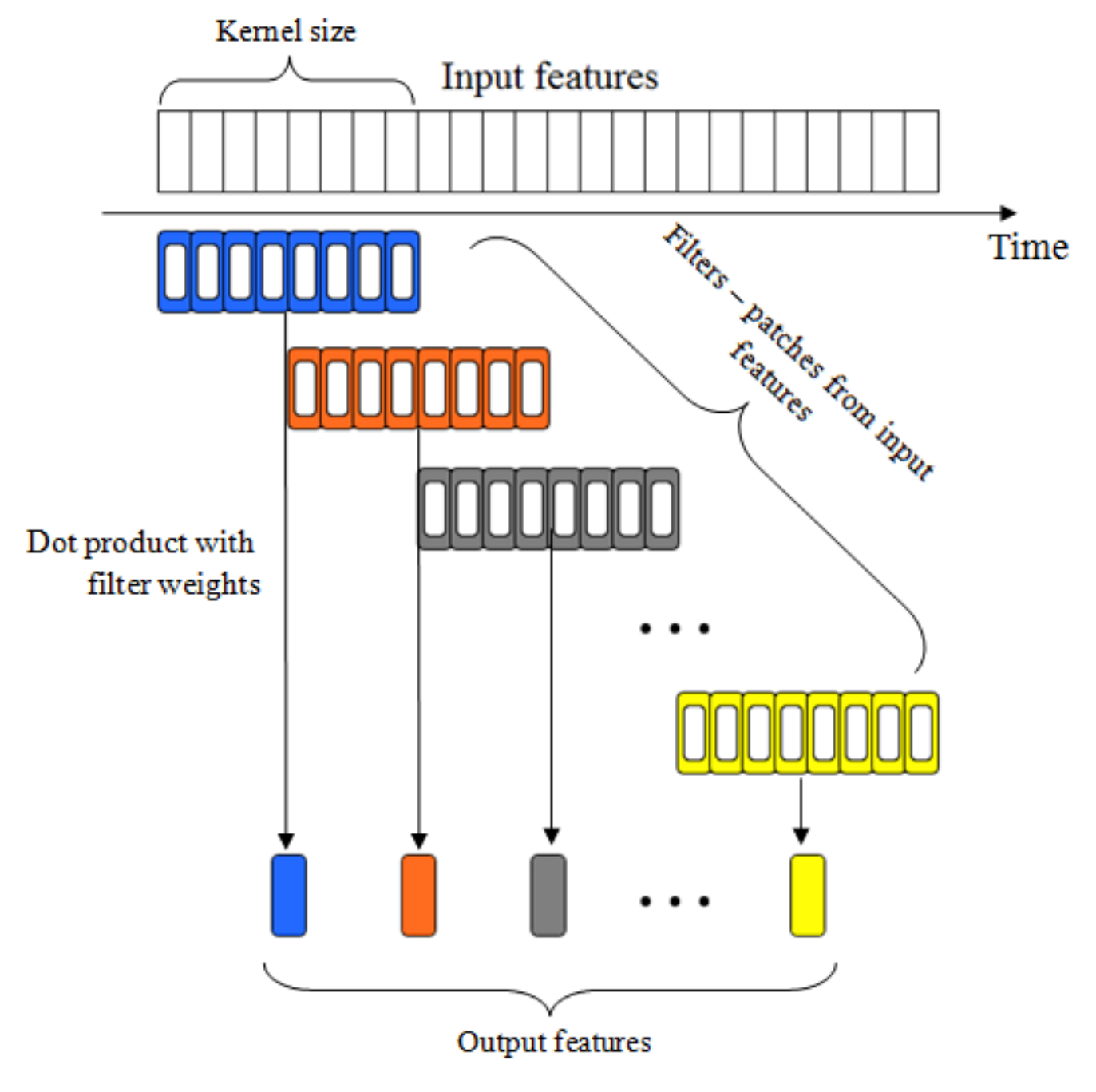}}
  \caption{Structure of 1D convolution layer.}
  \label{fig:1D-CNN}
\end{figure}

CNNs are DL architectures that achieved outstanding results in computer vision and image classification problems, due to their ability to extract features from local input patches through the application of relevant filters. CNNs can effectively learn the hierarchical features to construct a final feature set of a high level abstraction, which are then used to form more complex patterns within higher layers \cite{b20}. The same ideas can be applied to 1D-sequences of data, where 1D-CNNs are proven to be effective in deriving features from fixed-length segments of the data set. This characteristic of the 1D-CNN, together with the fact that the 1D convolution layers are translation invariant, which means that a pattern learned at a certain position in the signal can be latter recognized at a different position (e.g., the start instant of the packet), makes this architecture suitable for packet detection task.

Two types of layers are applied in compact 1D-CNNs: i) 1D-CNN layer, where 1D convolution occurs, and ii) Fully Connected (FC) layer. Each hidden CNN layer performs a sequence of convolutions, whose sum is passed through the activation function \cite{b21}. The main advantage of 1D-CNN represents fusing feature extraction and classification operations into a single process that can be optimized to maximize the network performance, because CNN layers process the raw 1D data and extract features used by FC layers for prediction tasks (Fig. \ref{fig:1D-CNN}). As a consequence, low computational complexity is provided and if compared to 2D-CNNs, 1D-CNN can use larger filter and convolution window sizes since the only expensive operation is a sequence of 1D convolutions.

\subsubsection{Training procedure} To train DNN models, the mean-squared error (MSE) loss: $L_{MSE}(\tau_{S}, \hat{\tau}_{S}) = \sum_{i}(\tau_{S_i}-\hat{\tau}_{S_i})^2$ is minimized, which achieves better performances as compared to the mean-absolute error (MAE) and Huber loss functions. The training set is separated into mini-batches of size 80, and 400 epochs are sufficient for the loss function convergence. In order to optimize network parameters, stochastic gradient descent (SGD) with Adam at the learning rate $\alpha = 0.001 $, $\beta_{1} = 0.9$ and $\beta_2 = 0.999$ is used \cite{18}.

\begin{table}[tbhp]
\caption{ 1D-CNN network parameters for packet detection.}
\begin{center}
\begin{tabular}{|c|c|}
\hline
Layer&Size (number of filters/neurons) \\
\hline 
Conv1D + ReLU&9\\
Conv1D + ReLU&5 (filter size is 3 samples)\\
FC + ReLU&3\\
Output (Linear)&1\\
\hline
\end{tabular}
\label{table_2}
\end{center}
\end{table}

The same 1D-CNN architecture (Table \ref{table_2}) is used for all experiments. Filter size of the first convolution layer is chosen as a half of the STS sample-length (8 samples), and stride of 1 sample is applied (Fig. \ref{fig:1D-CNN}). Note that we do not exploit full flexibility of 1D-CNN architecture since we apply a fixed number of input channels as well as the fixed-length filters. We apply such fixed architecture to make the analysis of the proposed algorithm in terms of its  performance and complexity easier. We note that the further optimization of the number of input channels and the input filter lengths may further improve performance vs complexity trade-off.

\subsection{Data Set Generation}  
\label{Data_set_gen}

\subsubsection{Simulated environment} The data set consists of ($|\pmb{y}|, \tau_{S}$) pairs, where $\tau_{S}$ indicates a packet start sample inside the block. Within the data set, we included about $50 \%$ of the blocks that do not contain a packet start instance, tagged with the value of $\tau_{S}=-1$. Among such blocks, roughly half contain only noise samples, while the other half contain intermediate or tail-parts of NDP packets. For data set blocks containing packet start instants $\tau_{S}$, its value is set uniformly at random among the input block samples. Data sets are created for input blocks $|\pmb{y}|$ of lengths: 40, 80, 160, 320, 800, 1600 samples, where the number of received blocks in each data set is 50000. Note that the larger the length of the input block, the complexity of the first layer increases, however, the number of blocks to be processed per unit time decreases. Careful complexity analysis is presented in Sec. \ref{Numerical_res}. From the data set, $70\%$ records are used for training, $15\%$ for validation and $15\%$ for testing.

Regardless of the input block size, all packets are simulated under the same conditions using the standard-compliant IEEE 802.11ah physical layer simulator. In order to examine estimator robustness to varying signal-to-noise-ratio (SNR), SNR values are uniformly and randomly selected from range [$1~dB$, $25~dB$]. During the simulations, indoor multipath fading channel - model B \cite{b17} is applied.

\subsubsection{Real-World Environment} 
\begin{figure}[htbp]
 \centerline{\includegraphics[width=1\columnwidth, height=2in]{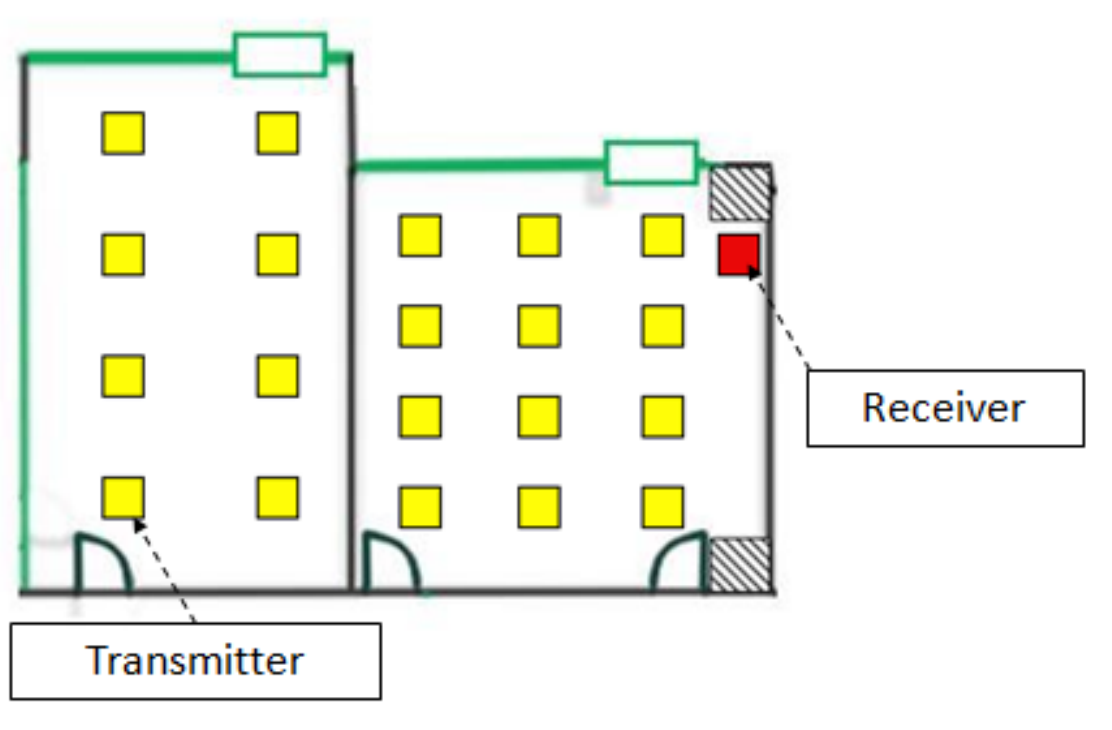}}
  \caption{Real world setup - transmitter and receiver positions.}
  \label{fig:Real_world_setup}
\end{figure}

In order to evaluate the proposed method in a real-world environment, we collect data sets using Software-Defined Radio (SDR) implementation. We deploy our real-world setup in an indoor environment, placing the transmitter along a sequence of predefined grid points, while the receiver is stationary, as shown in Fig. \ref{fig:Real_world_setup}. Note that 12 out of 20 transmitter positions are in the same room as the receiver, while the remaining 8 are in the neighboring room, thus providing us with the data set of a wider range of received SNRs.

\begin{figure}[htbp]
 \centerline{\includegraphics[width=0.9\columnwidth, height=3.5in]{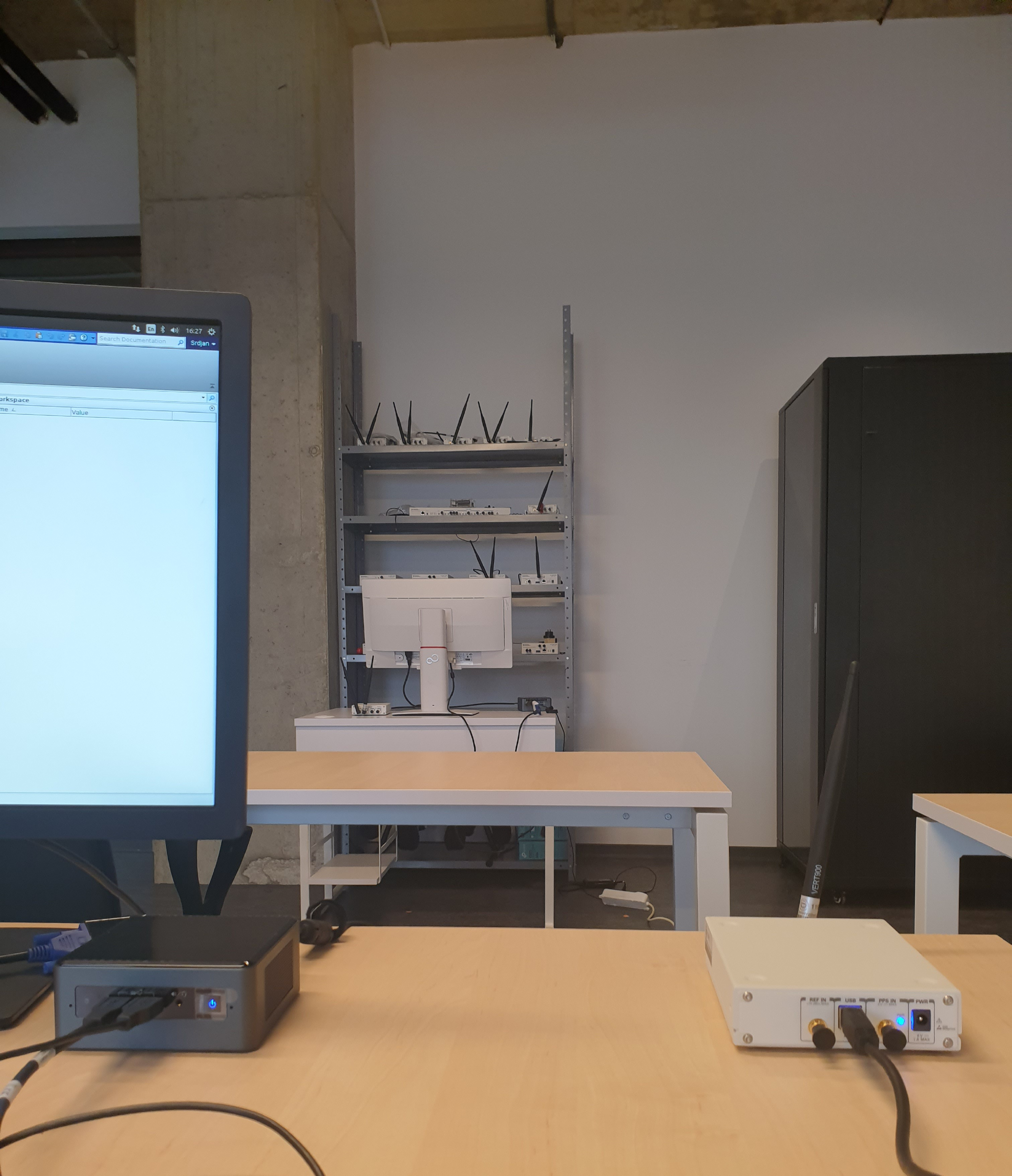}}
  \caption{Real world setup - USRP B210 SDR platforms}
  \label{fig:Real_world_setup_1}
\end{figure}

 Both the transmitter and the receiver include standard-compliant MATLAB-based 802.11ah PHY implementation and USRP B210 SDR platforms, as shown in Fig. \ref{fig:Real_world_setup_1}. From each point, the transmitter sends 1000 1 MHz NDP packets with the measured SNR range $\in [-6~dB, 31~dB]$). At the receiver side, the complex baseband data samples obtained after filtering and downsampling are collected and separated into the input blocks $|\pmb{y}|$ of lengths: 40, 80, 160, 320, 800 and 1600 samples. Roughly 50\%\ of blocks that do not contain packet start instance are included, resulting in a data set that consists of 40,000 $(|\pmb{y}|, \tau_S)$ pairs (70\%\ for training, 15\%\ for validation and 15\%\ for testing). Other system assumptions and parameters are the same as in the simulated environment.  
 
\subsection{Numerical Results}
\label{Numerical_res}

In this subsection, we discuss the packet detection performance of both CNN-based and conventional methods in terms of the mean absolute error (MAE) under different SNRs in the simulated environment. In the real-world environment, we present MAE averaged across the entire SNR range. Also, miss detection and false alarm rates are investigated and taken into account. Furthermore, we investigate the computational complexity of the proposed CNN-based algorithm for packet detection for different input block lengths, and compare them to the conventional method in terms of the approximate number of floating point operations per second (FLOPS). 

The complexity of the DL-based algorithms considered in this paper are evaluated for an inference phase only. In other words, we assume that the training process is done offline. Note that the offline training can be made more efficient by first pretraining the model on a realistic system simulator, and then extending the training with an additional, usually smaller, set of training samples collected from a real-world environment \cite{Ye_2018}. This process can be further improved by techniques of deep transfer learning, which can speed up the model design, as suggested in \cite{Alves_2021}. Also, authors in \cite{Elbir} propose effective combining of the trained models using the concept of federated learning in order to arrive at more robust and efficient models.

 \subsubsection{Simulated Environment}

\begin{figure}[htbp]
 \centerline{\includegraphics[width=1\columnwidth, height=0.9\columnwidth]{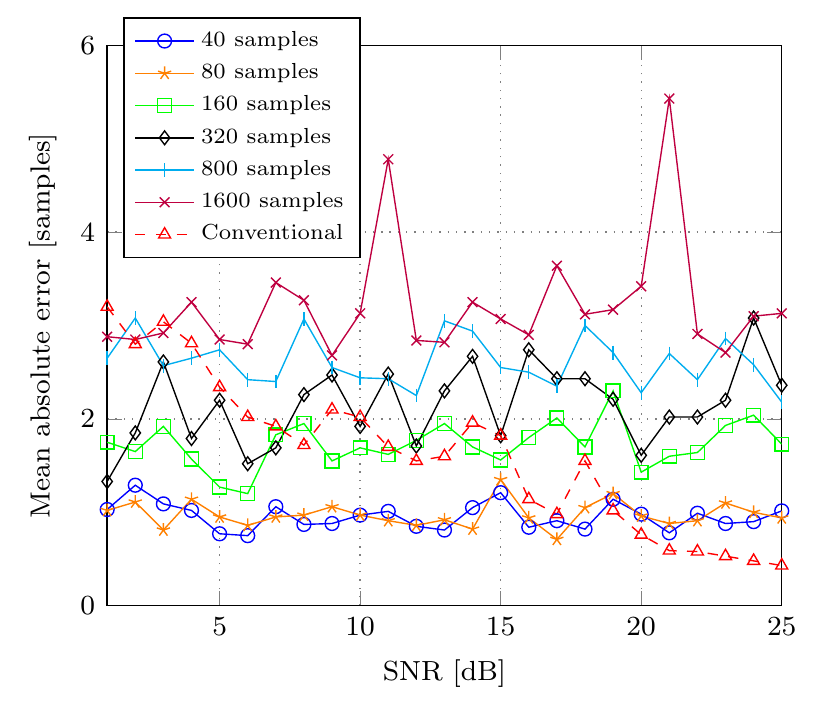}}
  \caption{MAE performance of 1D-CNN vs conventional packet detection for different received SNR.}
  \label{fig:MAE_O}
\end{figure}

In the simulated experiments, the number of 1D-CNN input channels is set to 4 for all input block lengths. Fig. \ref{fig:MAE_O} presents MAE packet detection performance of 1D-CNN architectures as a function of the received SNR evaluated over the test set. The figure also includes the results obtained by using conventional method after both coarse and fine packet start sample estimation is applied. We note that the 1D-CNN approach demonstrates better robustness to the variations of SNR as compared to the conventional method that deteriorates at lower SNRs. In addition, as the input block lengths decrease, the 1D-CNN packet detector outperforms the conventional method. Although this can be attributed to the fact that the estimated packet start sample value $\tau_{S}$ is bounded by the input block size (thus the estimation error naturally reduces by decreasing the input block length), we still note that 1D-CNN processing input blocks as large as 320 samples performs comparably with the conventional detector that slides across input blocks of 80 samples (Sec. \ref{Conventional_packet_det}), while outperforming the conventional detector for SNRs below 7 dB. Finally, it is interesting to compare the performance of different algorithms at SNR equal to 10 dB, since the authors in \cite{Handover} emphasize this SNR value as critical for different IEEE 802.11ah use cases. From Fig. \ref{fig:MAE_O} we note that the conventional algorithm has comparable performances with the CNN-based algorithm for an input block of 320 samples, while for the smaller input blocks, the CNN-based algorithm outperforms the conventional one.

\begin{figure}[htbp]
 \centerline{\includegraphics[width=1\columnwidth, height=0.9\columnwidth]{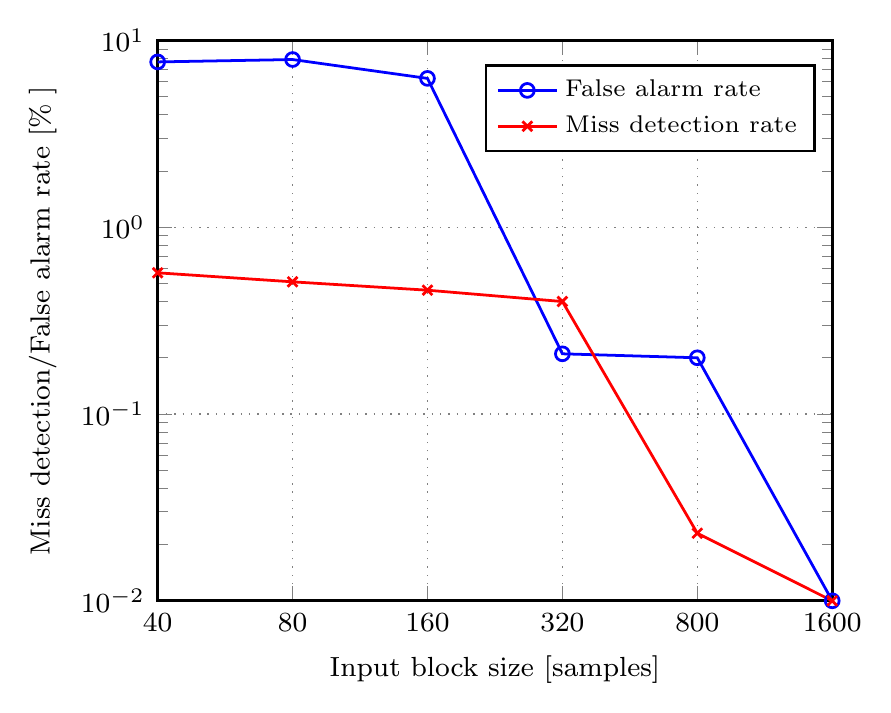}}
  \caption{1D-CNN packet detection miss detection and false alarm rates for different input block sizes.}
  \label{fig:MD/FD_O}
\end{figure}

For the same setup, Fig. \ref{fig:MD/FD_O} presents the miss detection and false alarm rates for different input block sizes. The results are expressed as a percentage of miss or false detected packets averaged across the entire test set (i.e., across all SNRs). For comparison, for the same testing conditions, the conventional method exhibits superb performance of miss detection rate equal $0.0012 \%$ and false alarm rate equal $0.0016 \%$. For 1D-CNN-based packet detectors, although the results vary across the range of input block lengths showing particularly high false alarm rates for small input block sizes, the performance gradually improves for larger input block lengths, achieving sub-$0.1\%$ miss detection and false alarm rates.

\subsubsection{Real-World Environment} Next, we explore the performance of the 1D-CNN-based packet detector in the real-world environment. The number of 1D-CNN input channels is kept at 4 for all input block lengths. Note that, in the simulated environment, the test data set contains approximately the same number of packets at each SNR value, thus we present MAE performance as a function of the received SNR (Fig. \ref{fig:MAE_O}). However, in a real-world environment, we do not have such control over received SNRs, and our data set is highly irregular in terms of recorded received SNR values. For this reason, average MAE across the whole range of SNR values is presented for each input block size, along with the performance of the conventional algorithm included as a benchmark. 

\begin{figure}[htbp]
 \centerline{\includegraphics[width=1\columnwidth, height=0.9\columnwidth]{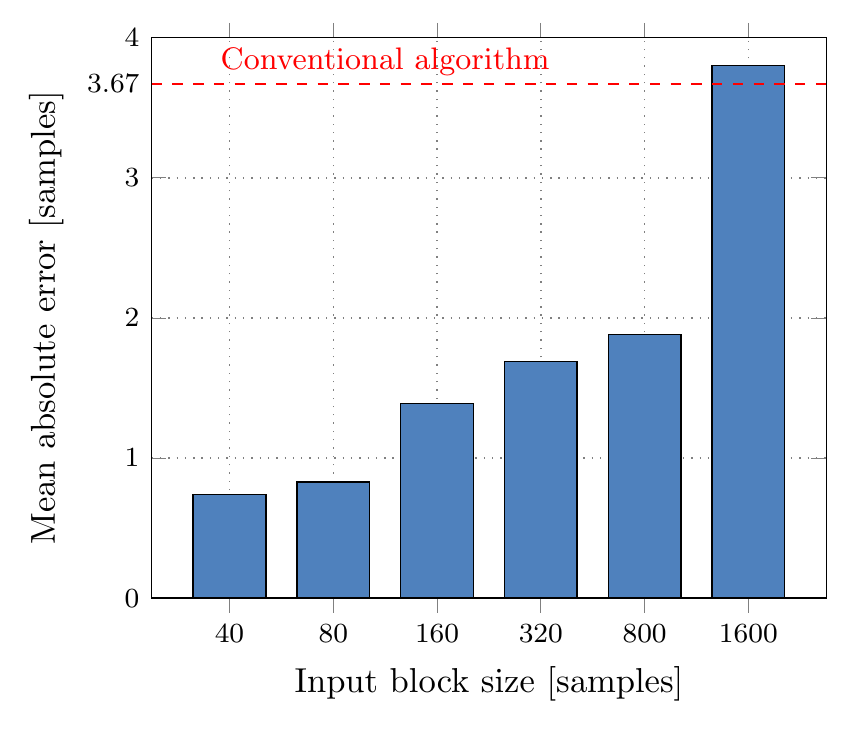}}
  \caption{ 1D-CNN MAE performance for different input block sizes in the real-world environment.}
  \label{fig:MAE_real_world}
\end{figure}

Fig. \ref{fig:MAE_real_world} shows that the proposed CNN-based algorithm outperforms the conventional method in terms of the averaged MAE. Moreover, such performance is achieved for input block lengths up to 800 samples, while for the input block length of 1600 samples, the performances of the two methods are similar.

\begin{figure}[htbp]
 \centerline{\includegraphics[width=1\columnwidth, height=0.9\columnwidth]{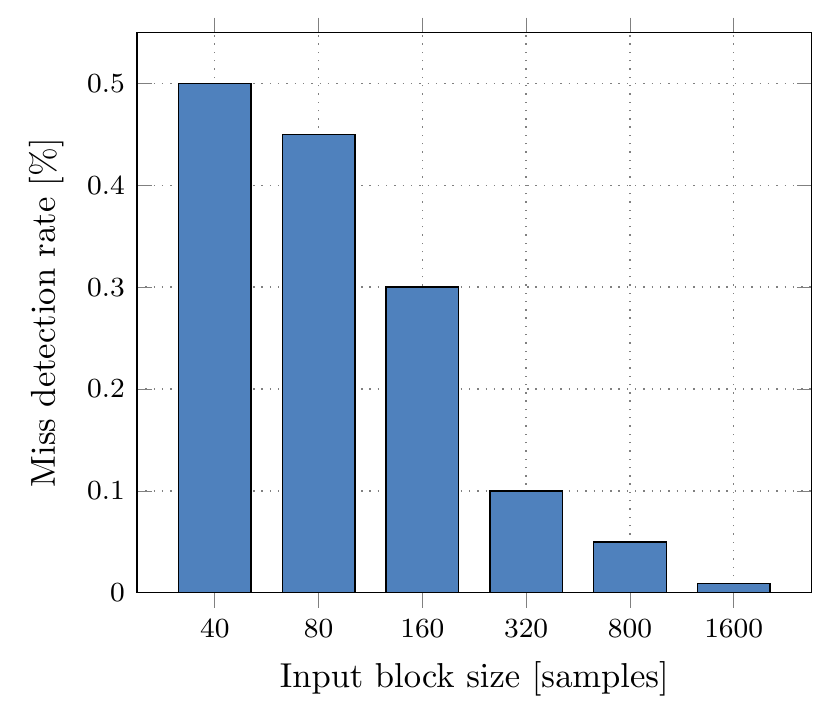}}
  \caption{ 1D-CNN miss detection rate for different input block lengths in the real-world environment.}
  \label{fig:MD_real_world}
\end{figure}

Fig. \ref{fig:MD_real_world} shows probability of miss detection for different input block sizes averaged across all SNRs. The obtained rates are promising as, even in the worst-case input block size of 40 samples, the obtained rates are below 0.5\%. For larger block lengths, the miss detection rates drop significantly, reaching as low as 0.01\%\ for input block size of 1600 samples. The conventional method is an order of magnitude better achieving the miss detection rate of 0.0026\%. 

\begin{figure}[tbp]
 \centerline{\includegraphics[width=1\columnwidth, height=0.9\columnwidth]{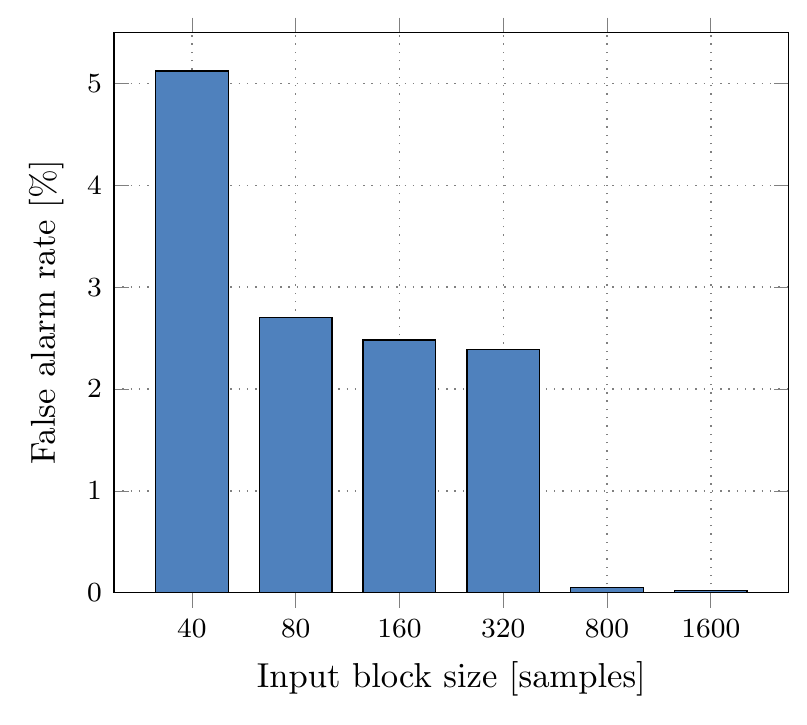}}
  \caption{ 1D-CNN false alarm rates for different input block lengths in the real-world environment.}
  \label{fig:FD_real_world}
\end{figure}

In terms of the false alarm rate, the proposed CNN-based algorithm for packet detection shows deteriorated performances. From Fig. \ref{fig:FD_real_world}, one can note that for small input block sizes, the false alarm rate can be as high as 5\%\, while with the increase in the input block length, the false alarm performance improves. The best achieved false alarm rate for  the CNN-based estimator of 0.015\%\ (input block length of 1600 samples) still falls short of the conventional algorithm whose false alarm rate is 0.0027\%. Finally, we note that the performance trends observed in the simulated environment are preserved in the real-world environment.

\subsubsection{Computational Complexity Analysis} 
\begin{figure}[htbp]
 \centerline{\includegraphics[width=1\columnwidth, height=0.9\columnwidth]{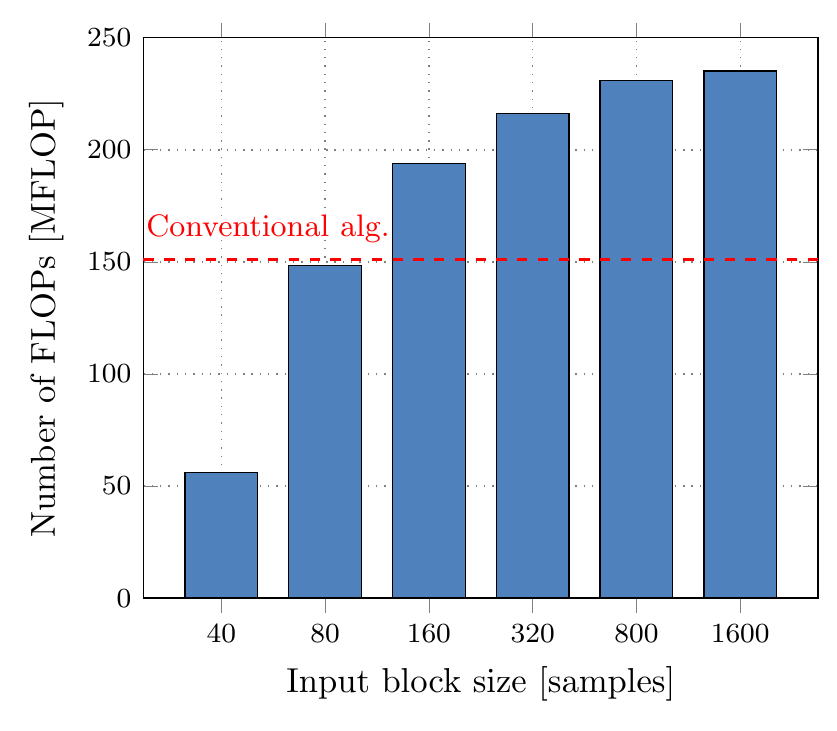}}
  \caption{Computation complexity comparison for 1D-CNNs and conventional packet detectors.}
  \label{fig:FlopBar}
\end{figure}

Assuming the sampling rate of 1 MHz for IEEE 802.11ah scenario used in our experiments, Fig. \ref{fig:FlopBar} shows the approximate number of FLOPS of the 1D-CNN architecture as a function of the input block lengths, with the conventional method included for reference. The complexity of each layer of 1D-CNN may be computed by calculating the number of additions and multiplications within each layer. The total number of FLOPS for a CNN depends on the input block size, however, note that although larger input blocks lead to more complex network, they also reduce the number of blocks processed per second. According to \cite{Karra_2017}, the complexity of a single convolution layer depends on filter length $F$, number of input ($ch_i$) and output ($ch_o$) channels, and output width $K$, while the complexity of FC layer is determined by the input ($N_i$) and the output ($N_o$) size. Mathematical expressions used for calculating an approximate number of FLOPS (multiplications and additions) in a single layer are given in Table \ref{table_4} \cite{Karra_2017}. 

\begin{table}[tbp]
\caption{1D-CNN and FC Approximate layer complexity }
\begin{center}
\begin{tabular}{|c|c|}
\hline
Layer/Operation&Expression  \\
\hline 
Conv1D/ MUL&$F\ast ch_i \ast ch_o \ast K$\\

Conv1D/ADD&$F\ast (ch_i+1) \ast ch_o \ast K$\\

FC/MUL&$N_i \ast N_o$\\

FC/ADD&$(N_i+1) \ast N_o$\\
\hline
\end{tabular}
\label{table_4}
\end{center}
\end{table}

Regarding the conventional method, it consists of two parts: coarse and fine estimation. During the coarse estimation, it uses sample-by-sample processing of input blocks of length 80 samples. The FLOPS count for the coarse packet detection is derived by calculating the number of multiplications and additions for a single input block of length 80 samples, multiplied by the number of blocks processed per second. Complexity of the fine estimation, which is run only when the coarse estimation detects the start of the packet, is neglected.

For smaller input block lengths, Fig. \ref{fig:FlopBar} shows that the complexity of 1D-CNNs is lower or comparable with the conventional algorithm. Taking the overall results into account, 1D-CNN offers a relatively wide operational range for balancing between MAE, computational complexity in MFLOPs, miss detection and false alarm rates. We summarize our findings as follows: 1D-CNNs are able to outperform conventional methods under reduced computational effort, while being inferior in miss detection and false alarm rates.

\section{Preamble-Based CFO Estimation}

In the second part of the paper, we consider implementation of deep-learning based CFO estimation in IEEE 802.11ah and compare its performance with the conventional method.

\subsection{Conventional CFO Estimation}
\label{Conv_CFO}

A common approach to CFO estimation uses the fact that the samples of two consecutive identical short training symbols differ by the phase shift proportional to the CFO $f_{off}$:

\begin{equation}
    y_{\tau+L} = y_{\tau} e^{j 2 \pi f_{off} T_s},
\end{equation}
where $T_s$ represents the sample period \cite{b9}. Maximum likelihood CFO estimate uses the phase of complex correlation $\Lambda_\tau$ (Eq. 4) between the repeated training symbols,  denoted as $\hat{\phi} = \angle(\Lambda_\tau)$, to estimate CFO \cite{b3}, \cite{b10}:

\begin{equation}
    \hat{f}_{off}=\frac{f_s\hat{\phi}}{2\pi N},
\end{equation}
where $f_s=\frac{N}{T_s}$ is the sample frequency. 

 In the IEEE 802.11ah scenario, the CFO estimation can be separated into two steps. The coarse CFO, denoted as $\hat{f}^{(1)}_{off}$, is carried out using auto-correlation of two adjacent STS within STF, taken at the estimated packet start sample time $\tau_S$ \cite{b4}:

\begin{equation}
    \Lambda_{\tau_S}^{(1)}=\sum_{n=\tau_S}^{\tau_S+P-l_S} y^{*}_n y_{n+l_{S}} = e^{\frac{j2\pi \hat{f}^{(1)}_{off} l_{S}}{f_s}} \sum_{n=\tau_S}^{\tau_S+P-l_S} |y_n|^2
\end{equation}
where  $P$ is equal to or is a multiple of $l_{S}$. Using (10) and (11), and $\hat{\phi}^{(1)} = \angle(\Lambda_{\tau_S}^{(1)})$, we get:

\begin{equation}
    \hat{f}^{(1)}_{off}=\frac{f_s}{2\pi l_{S}} \hat{\phi}^{(1)}.
\end{equation}

After correcting $\hat{f}^{(1)}_{off}$ over the signal $\pmb{y}$, the coarse CFO-compensated signal $\hat{\pmb{y}}$ is obtained. Using LTF field of $\hat{\pmb{y}}$, the fine CFO estimation $\hat{f}^{(2)}_{off}$ can be expressed as \cite{b4}:    
\begin{equation}
    \Lambda_{\tau_L}^{(2)}=\sum_{n=\tau_L}^{\tau_L+L_L-l_L} \hat{y}^{*}_{n}\hat{y}_{n+l_{L}} = e^{\frac{j2\pi \hat{f}^{(2)}_{off} l_{L}}{f_s}} \sum_{n=\tau_L}^{\tau_L+L_L-l_L} |\hat{y}_n|^2,
\end{equation}
where $\tau_{L}=\tau_{S}+L_{S}$ is the initial LTF sample, $L_{L}$ is a sample-length of LTF field, and $l_{L}$ is a sample-length of a long training symbol. Using $\hat{\phi}^{(2)} = \angle(\Lambda_{\tau_L}^{(2)})$ the fine CFO is estimated as:

\begin{equation}
    \hat{f}^{(2)}_{off}=\frac{f_s}{2\pi l_{L}}\hat{\phi}^{(2)}
\end{equation}

Finally, the CFO of the received signal is estimated as the sum of the coarse and fine CFOs: $\hat{f}_{off} = \hat{f}^{(1)}_{off} + \hat{f}^{(2)}_{off}$.

\subsection{Deep-Learning Based CFO Estimation}
\label{DL_cfo}

 In this paper, we test the ability of selected DNN architectures to estimate the CFO from the phase of received STF samples:
\begin{equation}
    \hat{f}_{off} = f(\angle(\pmb{y}_{STF})).
\end{equation}
In other words, a DNN architecture learns the mapping between the received $\angle(\pmb{y}_{STF})$ and $f_{off}$. Note that we test the DNN-based CFO estimation only on the STF field, unlike the conventional methods that use both STF and LTF fields. Finally, we note that in both simulation and real-world experiments in this paper, the CFO estimation is applied sequentially after the conventional packet detection is applied. Thus the effects of imperfect packet detection are included in the CFO estimation results in Sec. \ref{CFOres}.
Next, we detail the DNN architectures considered for CFO estimation, and describe the data set and training procedure.

\subsubsection{Fully Connected Feed-Forward Neural Networks} This neural network architecture consists of an input, an output and the set of hidden layers, and is a simple and well-understood DNN model. The relation between the input $\textbf{x}$ and the output $\textbf{y}$ is a layer-wise composition of computational units:

\begin{equation}
     \textbf{y} = f(\textbf{x}, \mathbf{\Theta}) = f_o(g_{M-1}(f_{M-1}(\ldots(g_1(f_1(\textbf{x})))))), 
\end{equation}
where $\mathbf{\Theta}$ denotes the set of network parameters: weights $\mathbf{W_i}$ and biases $\mathbf{b_i}$, $f_i(\textbf{x}) = \mathbf{W_i}\textbf{x}+\mathbf{b_i}$ and $g_i(\cdot)$ are the linear pre-activation and activation function of the $i_{th}$ hidden layer, respectively, $f_o(\cdot)$ represents the linear function of the output layer, and $M$ is the number of layers. Among the non-linear activation functions, we focus on rectified linear units (ReLU), as ReLU DNNs are known universal piece-wise linear function approximators for a large class of functions \cite{b11}. 


\subsubsection{Recurrent Neural Networks} RNNs represent sequence-based  models able to establish temporal correlations between the previous and the current circumstances. As such, RNN represent a suitable solution for the CFO estimation problem, given that the estimated CFO values between the samples of the subsequent symbols in the past have influence on the current CFO estimate. 

\begin{figure}[htbp]
 \centerline{\includegraphics[width=1\columnwidth, height=1.8in]{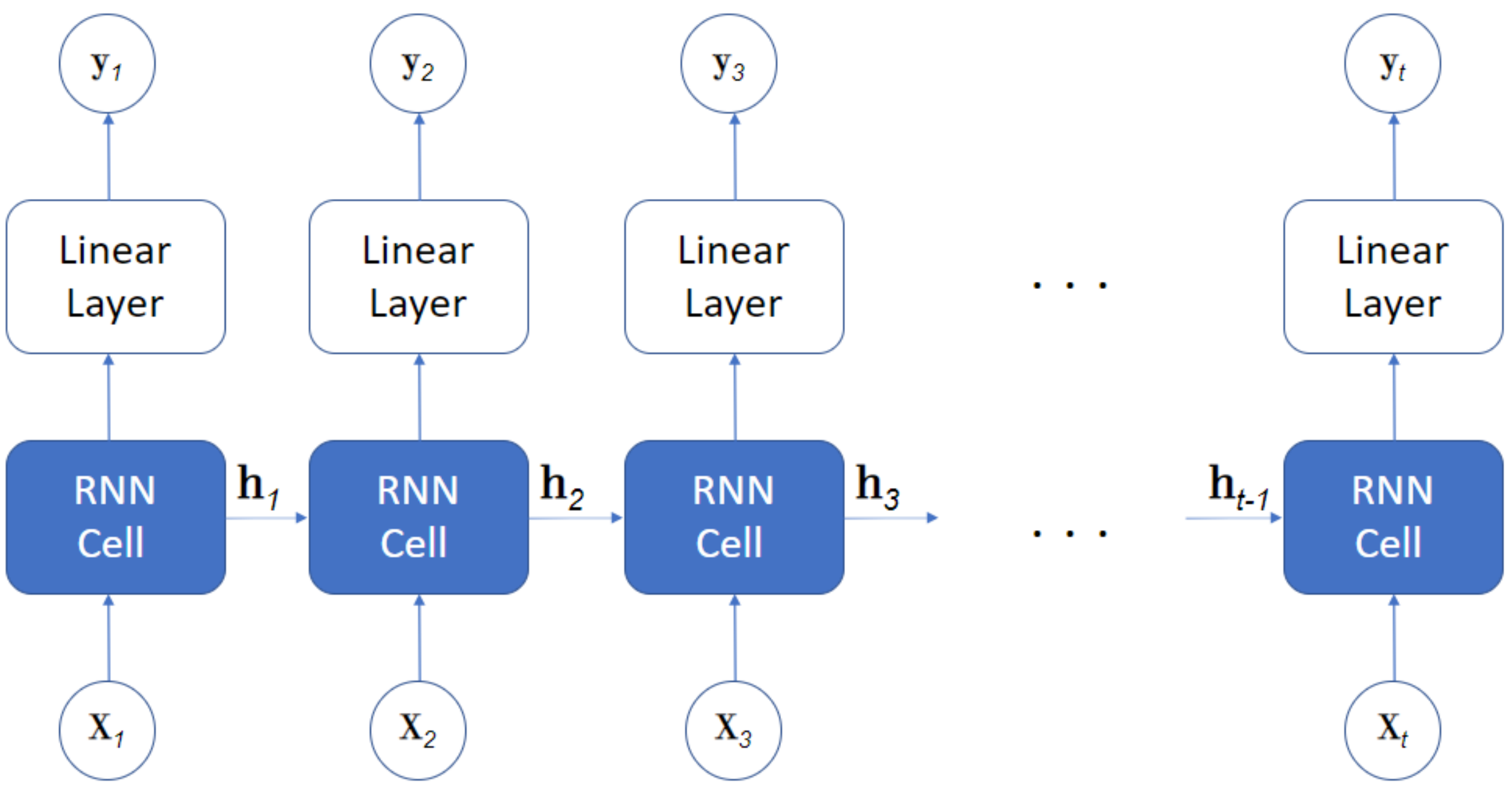}}
  \caption{The structure of Recurrent Neural Network.}
  \label{fig:RNN}
\end{figure}

A simple example of a single-layer RNN is given in Fig. \ref{fig:RNN}, where the output of the previous time step $t-1$ becomes a part of the input of the current time step $t$, thus capturing past information. Computation result performed by one RNN cell can be expressed as a following function \cite{b12}:
\begin{equation}
    \mathbf{h_t} = \tanh(\mathbf{W}_{ih}\mathbf{x}_t + \mathbf{b}_{ih} + \mathbf{W}_{hh}\mathbf{h}_{t-1} + \mathbf{b}_{hh}),
\end{equation}
where $\tanh$ represents the hyperbolic tangent function, $\mathbf{h}_t$ and $\mathbf{h}_{t-1}$ are the hidden states at time steps $t$ and $t-1$, respectively, $\mathbf{W}_{ih}$, $\mathbf{W}_{hh}$ and $\mathbf{b}_{ih}$, $\mathbf{b}_{hh}$ are the weights and the biases which need to be learned, and an input at time $t$ is denoted as $\mathbf{x}_{t}$.

Basic RNN cells fail to learn long-range dependencies due to the vanishing or exploding gradients. To solve this, Long Short-Time Memory (LSTM) \cite{b13} cells are put forward that contain special units called \textit{memory blocks} in recurrent hidden layer, which enhance its capability to model long-term dependencies. This block is a recurrently connected subnet that contains functional modules called memory cells and gates. The former remembers the network temporal state while the latter controls the information flow from the previous cell state.


Besides standard LSTM cells, we also consider Gated Recurrent Unit (GRU) \cite{b15}. The main ideas from LSTMs are preserved, but GRU introduces only two gates, update gate and reset gate, to control the information flow. GRUs perform similarly to LSTM, but with reduced execution time \cite{b16}.

\subsubsection{Training Procedure} To train DNN models, we minimize the MSE loss: $L_{MSE}(f_{off}, \hat{f}_{off}) = \sum_{i}(f_{off_i}-\hat{f}_{off_i})^2$. The training set is divided into mini-batches of size 100, and 500 epochs are sufficient for the loss function convergence. Network parameters are optimized the same way as in Sec. \ref{DL_packet_det}, i.e., by using SGD with Adam at the learning rate $\alpha = 0.001 $, $\beta_{1} = 0.9$ and $\beta_2 = 0.999$ \cite{18}.





The parameters of the proposed ReLU DNN and RNN architectures are described in Table \ref{tableFNN} and Table \ref{table_RNN_CFO}, respectively. Unlike ReLU DNN, where the input is the whole sequence $\angle(\pmb{y}_{STF})$, at RNN, this sequence is split into STSs (16 samples), and one STS is input into one LSTM/GRU unit. 

\begin{table}[tbhp]
\caption{ ReLU DNN network parameters for CFO estimation.}
\begin{center}
\begin{tabular}{|c|c|}
\hline
Layer&Size (number of neurons) \\
\hline 
Input (Linear)&160\\
FC + ReLU&32\\
FC + ReLU&64\\
FC+ ReLU&16\\
Output (Linear)&1\\
\hline
\end{tabular}
\label{tableFNN}
\end{center}
\end{table}

\begin{table}[tbhp]
\caption{RNN network parameters for CFO estimation.}
\begin{center}
\begin{tabular}{|c|c|}
\hline
Layer&Size (number of units/neurons) \\
\hline 
LSTM/GRU&30\\
FC + ReLU&5\\
Output (Linear)&1\\
\hline
\end{tabular}
\label{table_RNN_CFO}
\end{center}
\end{table}

\subsection{Data Set Generation}

\subsubsection{Simulated Environment} Using the simulated environment, we generate the data set of pairs ($\angle(\pmb{y}_{STF}), f_{off}$), where $f_{off}$ represents a CFO introduced during transmission. After downsampling and filtering, $\pmb{y}_{STF}$ consists of 160 samples (10 repetitions of 16-sample STS). We simulated transmission of $50,000$ NDP packets and extracted STF phase vectors, while the corresponding true CFO values are generated within the simulation uniformly at random from $[-\frac{\Delta f}{2}, \frac{\Delta f}{2}] = [-15.625~kHz, 15.625~kHz] $. From the data set, 70\% of the records are used for training, 15\% for validation and 15\% for testing purposes. In order to examine estimator robustness, NDP packets are received with different SNRs ranging between 1 $dB$ and 25 $dB$. Depending on the simulated channel model, two data sets are created: i) AWGN channel, and ii) indoor multipath fading channel - model B \cite{b17}.


\subsubsection{Real World Environment} The setup used for data set generation in the real-world environment is the same as in Sec. \ref{Data_set_gen}. From each grid point, the transmitter sends 1000 1 MHz NDP packets with the measured SNR range $\in [-6~dB, 31~dB]$. At the receiver side, after the packet detection, the STF phase vectors ($\angle(\pmb{y}_{STF})$) are extracted. The collected data set consists of 20,000 $(\angle(\pmb{y}_{STF}), \hat{f}_{off})$ pairs (70\%\ for training, 15\%\ for validation, and 15\%\ for testing), where as a label $\hat{f}_{off}$ we use a CFO estimated using the conventional algorithm. This is due to the fact that, in the real-world conditions, we do not have a priori knowledge on CFO introduced during the transmission. Thus, in this case, we train the DL-based CFO estimator to replicate the conventional method performance. Note also that, in contrast to the simulated environment where the CFO values are generated uniformly at random from a given interval, in real-world experiments, estimated CFO values between two SDR devices are nearly stationary.

\subsection{Numerical Results}
\label{CFOres}

In this subsection, the performance of the DL-based method is compared with the conventional one in both simulated and real-world environments. In addition, we compare the two methods in terms of the computation complexity evaluated using the approximate number of FLOPs per packet. As it is described in Sec. \ref{Numerical_res}, for the CFO estimation training is again done offline, so complexity analysis for DL-based algorithms is conducted only for the inference phase.  

\subsubsection{CFO Estimation Performance in Simulated Environment} 
\begin{figure}[htbp]
 \centerline{\includegraphics[width=1\columnwidth, height=0.85\columnwidth]{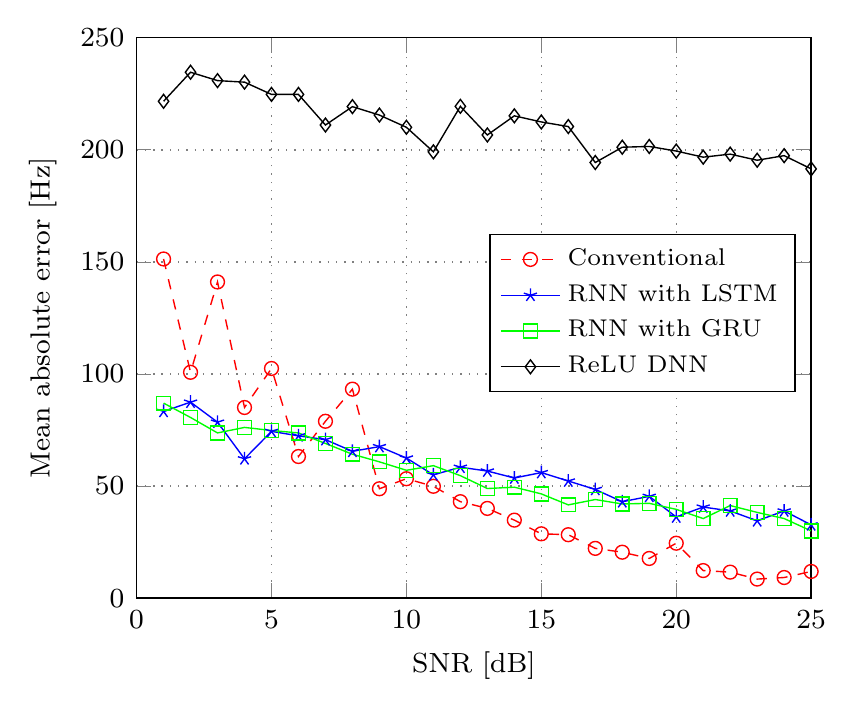}}
  \caption{MAE performance of different CFO algorithms for different received SNRs under AWGN channel.}
  \label{fig:CFO-MAE-AWGN}
\end{figure}

\begin{figure}[htbp]
 \centerline{\includegraphics[width=1\columnwidth, height=0.9\columnwidth]{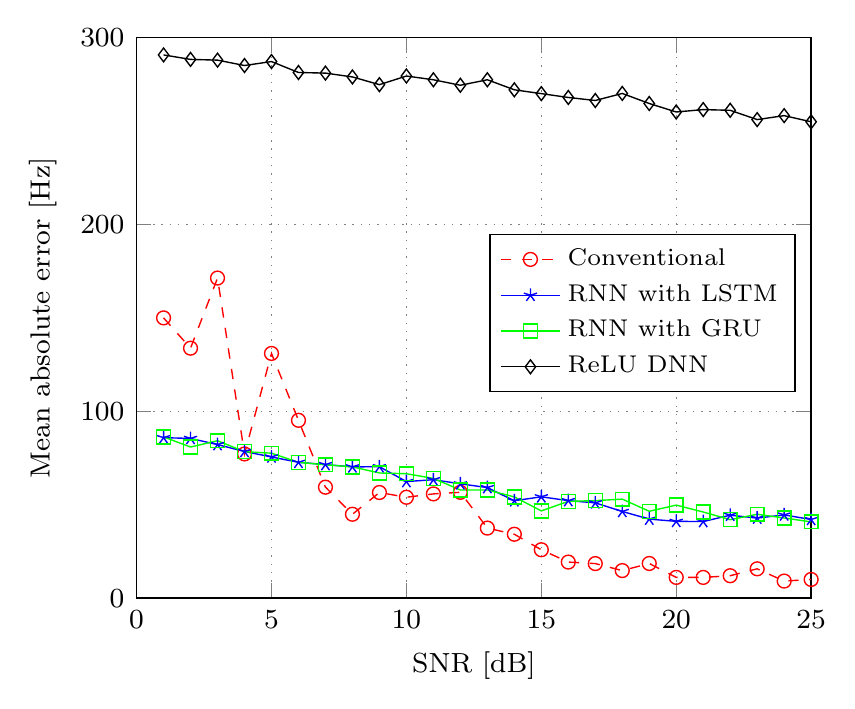}}
  \caption{MAE performance of different CFO algorithms for different received SNRs under indoor channel model.}
  \label{fig:CFO-MAE-ModB}
\end{figure}

MAE of CFO estimation as a function of channel SNR is presented in Figs. \ref{fig:CFO-MAE-AWGN} and \ref{fig:CFO-MAE-ModB} for both simulated channel models (see Sec. \ref{Data_set_gen}), respectively.

DNN-based methods use only STF samples as an input, while conventional methods use both STF+LTF samples through two-step coarse and fine CFO. We note that certain DNN approaches are more robust to varying SNR values than the conventional algorithm, which however outperforms all DNN architectures at the higher SNRs (above 8 dB). We also note that the more challenging indoor fading channel (model B) increases the MAE of all methods by approximately 15 Hz. As for the packet detection task, we observe that, for the SNR value of 10 dB, the conventional algorithm slightly outperforms the RNN-based method for both channel models.

\begin{figure}[htbp]
 \centerline{\includegraphics[width=1\columnwidth, height=0.9\columnwidth]{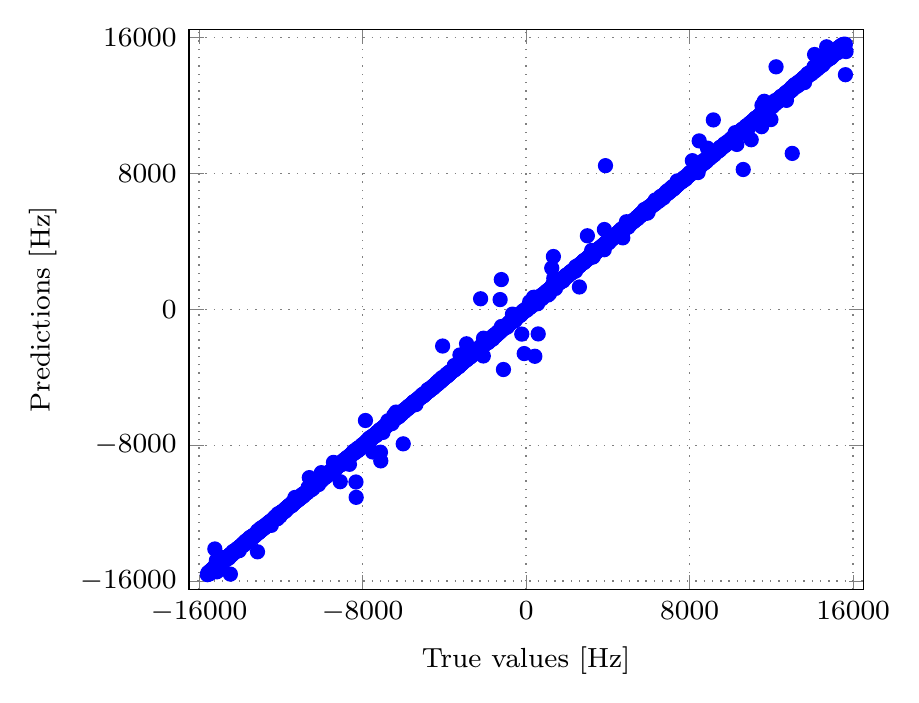}}
  \caption{True CFO values (x axis) vs CFO values predicted by RNN (y axis).}
  \label{fig: RNN outliers}
\end{figure}

We identify the existence of outliers as the main reason why RNN is not able to follow the MAE performance of the conventional method at high SNRs. Indeed, taking a closer look at Fig. \ref{fig: RNN outliers}, the majority of test samples are predicted with high accuracy, except a few that deviate and strongly affect MAE. In order to solve this problem, two different approaches are pursued: i) we extend the data set with additional 20000 samples, ii) we increase the RNN architecture complexity (using a single GRU layer with 50 units followed by a two ReLU FC layer with 30 and 20 neurons, respectively, and an output single-neuron layer). Our preliminary results demonstrate slight improvement only in the second approach, however, at high complexity costs (complexity will be discussed in Sec. \ref{CCAn}).  

The problem of outliers can be addressed by designing additional outlier detection methods. For example, one can include unsupervised methods such as deep autoencoders for outlier detection \cite{autoencoder}. We are currently investigating such methods, however, we note they will additionally contribute to the complexity of the proposed RNN-based method.

\subsubsection{CFO Estimation Performance in the Real World Environment} 
\begin{figure}[htbp]
 \centerline{\includegraphics[width=1\columnwidth, height=0.9\columnwidth]{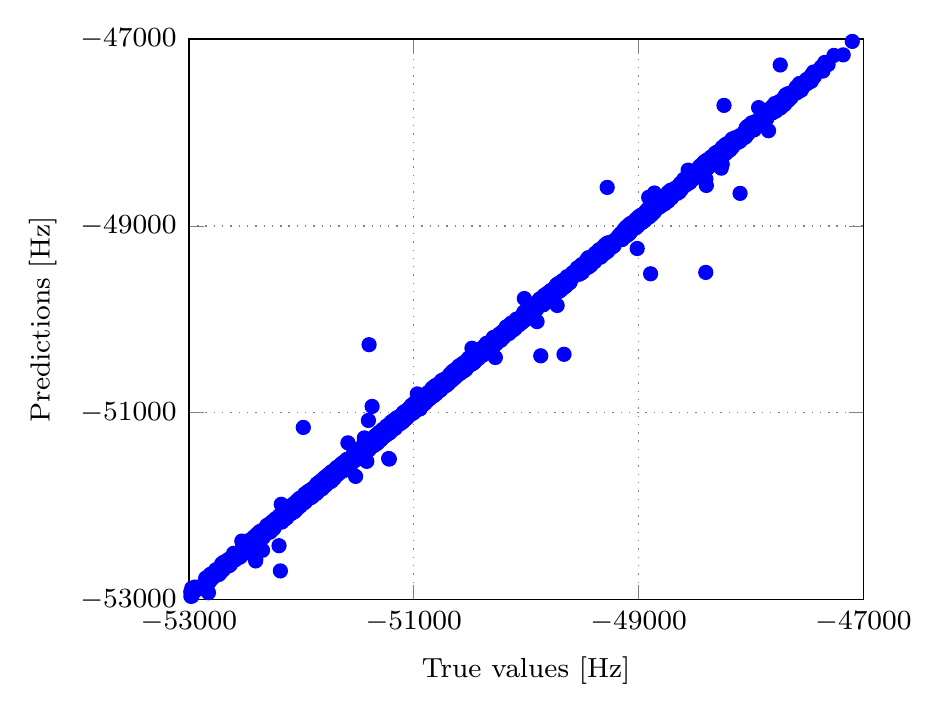}}
  \caption{CFO predicted by RNN vs estimated by conventional algorithm.}
  \label{fig:Real_lstm}
\end{figure}

We explore the ability of the proposed algorithm to replicate the results obtained using the conventional algorithm. Based on the MAE obtained in a simulated environment, we use RNNs with LSTM as a DNN-based method. Fig. \ref{fig:Real_lstm} shows that, except for a few outliers, the proposed RNN-based estimator is able to replicate the performance achievable with the conventional one.

\subsubsection{Computational Complexity Analysis}
\label{CCAn}
For the proposed RNN-based algorithm, an approximate number of FLOPs for processing a single packet is presented and compared with the complexity of the conventional algorithm in Table \ref{table_5}. The reason why we calculate the number of FLOPs per packet is due to the fact that the CFO estimation occurs only upon the packet detection event.

\begin{table}[tbhp]
\caption{RNN Approximate layer complexity }
\begin{center}
\begin{tabular}{|c|c|}
\hline
Layer/Operation&Expression  \\
\hline 
RNN/ MUL&$U^2+NF\ast U+2U$\\

RNN/ADD&$U^2+NF\ast U+U$\\
\hline
\end{tabular}
\label{table_6}
\end{center}
\end{table}

In order to calculate the number of FLOPs for the conventional algorithm, we take into account the number of multiplications and additions per packet for both coarse and fine CFO estimation. This number is favourable as, given the STF and LTF fields, the only task is to calculate the phase of the complex correlation as described in Sec. \ref{Conv_CFO}. On the other hand, the number of FLOPs for DNN-based algorithms is calculated as the number of multiplications and additions within each network layer. Since ReLU DNN comprises only  a FC layers, mathematical expressions for calculating the approximate number of FLOPs are described in Table \ref{table_4}. In Table \ref{table_6} \cite{nisar} we note that  the number of multiplications and additions in one recurrent cell depends on the number of recurrent units in a layer ($U$) and on  the number of features in one time stamp ($NF$, in our case we have 10 time stamps, each with 16 features). Except a recurrent layer, the proposed RNN also has a single FC layer whose complexity needs to be taken into account (Table \ref{table_4}) in order to obtain the total number of FLOPs. 

In Table \ref{table_6} we provide the expressions used to evaluate the computational complexity of a simple recurrent cell. In addition, LSTM or GRU units introduce additional memory cells and gates, having higher complexity than a simple recurrent cell. For example, the total number of FLOPs for a single LSTM cell is approximately 4 times higher than for a simple recurrent cell, while for a GRU cell, it is approximately 3 times higher than for a simple RNN cell. Finally, Table \ref{table_5} shows that, despite their excellent accuracy in terms of MAE, DNN-based methods suffer from high complexity in terms of the number of FLOPs per packet.

The complexity of the RNN architecture is the main reason why, instead of a single architecture, we used  different neural network architectures for packet detection and CFO estimation tasks. As our preliminary results show, when RNN is applied for packet detection task in the real-world environment (using the same parameters described in Table \ref{table_RNN_CFO}), MAE performances are slightly increased compared to 1D-CNN, i.e., they are comparable to the conventional algorithm, however, for the price of significant increase in the computation complexity.

\begin{table}[tbhp]
\caption{Approximate number of FLOPs for CFO estimation}
\begin{center}
\begin{tabular}{|c|c|}
\hline
Algorithm&Number of FLOPs  \\
\hline 
Conventional&\textbf{224}\\

LSTM&11651\\

GRU&8738\\

ReLU DNN&16209 \\
\hline
\end{tabular}
\label{table_5}
\end{center}
\end{table}

\section{Conclusion}

We performed an in-depth performance and complexity study of the DL-based packet detection and CFO estimation in preamble-based IEEE 802.11 systems. For both packet detection and CFO estimation, the conditions under which the performance of the DL-based methods approach or even surpass the conventional methods, but also, the conditions under which their  performance is inferior, are clearly presented. 

For the case of packet detection, 1D-CNNs are identified as the best-performing architecture able to achieve excellent accuracy that matches or even surpasses the conventional method (at low-to-medium SNRs), under favourable computation complexity. In contrast, the conventional method is always superior in terms of the false alarm and miss detection rate. For the case of CFO estimation, RNNs are identified as the best-performing architecture that are able to match the accuracy of the conventional method (at low-to-medium SNRs), however, their complexity is always inferior to conventional methods. Our findings are supported by numerical simulation results, and the real-world testbed using SDRs. According to our preliminary results for both packet detection and CFO estimation tasks, the proposed methods could be extended to other preamble-based IEEE 802.11 standards operating in 2.4/5 GHz bands.

Finally, for our future work, we plan to extend our investigation to multiple-input multiple-output (MIMO) modes of operation of IEEE 802.11ah standard, investigate effects of imperfect DL-based packet detection and CFO estimation on the DL-based channel estimation, and real-world implementation of the proposed methods in field-programmable gate array (FPGA) hardware in order to estimate realistic latency and resource requirements.

\end{document}